\documentclass[letterpaper]{article} 
\usepackage[table]{xcolor} 
\usepackage{aaai2026}  
\usepackage{times}  
\usepackage{helvet}  
\usepackage{courier}  
\usepackage[hyphens]{url}  
\usepackage{graphicx} 
\usepackage{booktabs}
\usepackage{multirow}
\usepackage{algpseudocode}
\usepackage{amssymb}
\usepackage{subfiles}
\usepackage{bibunits}      
\usepackage{natbib}        
\usepackage{caption} 
\usepackage{algorithm}
\usepackage{amsmath}
\usepackage{newfloat}
\usepackage{listings}
\DeclareCaptionStyle{ruled}{labelfont=normalfont,labelsep=colon,strut=off} 
\floatstyle{ruled}
\newfloat{listing}{tb}{lst}{}
\floatname{listing}{Listing}
\title{VoiceCloak: A Multi-Dimensional Defense Framework against Unauthorized Diffusion-based Voice Cloning 
}
\author{
    Qianyue Hu\textsuperscript{\rm 1},
    Junyan Wu\textsuperscript{\rm 1},
    Wei Lu\textsuperscript{\rm 1}\thanks{Corresponding author},
    Xiangyang Luo \textsuperscript{\rm 2}
}
\affiliations{
    \textsuperscript{\rm 1}School of Computer Science and Engineering, MoE Key Laboratory of Information Technology, Guangdong Province Key Laboratory of Information Security Technology, Sun Yat-sen University, Guangzhou 510006, China\\
    \textsuperscript{\rm 2}State Key Laboratory of Mathematical Engineering and Advanced Computing, Zhengzhou, 450002, China \\
    
    huqy56@mail2.sysu.edu.cn,
    wujy298@mail2.sysu.edu.cn,
    luwei3@mail.sysu.edu.cn,
    luoxy\_ieu@sina.com

}

\usepackage{bibentry}

\defaultbibliography{aaai2026}    
\defaultbibliographystyle{aaai2026}  

\begin{document}

\nocopyright

\begin{bibunit}
\maketitle
\begin{abstract}
Diffusion Models (DMs) have achieved remarkable success in realistic voice cloning (VC), while they also increase the risk of malicious misuse. 
Existing proactive defenses designed for traditional VC models aim to disrupt the forgery process, but they have been proven incompatible with DMs due to the intricate generative mechanisms of diffusion. 
To bridge this gap, we introduce VoiceCloak, a multi-dimensional proactive defense framework with the goal of obfuscating speaker identity and degrading perceptual quality in potential unauthorized VC. 
To achieve these goals, we conduct a focused analysis to identify specific vulnerabilities within DMs, allowing VoiceCloak to disrupt the cloning process by introducing adversarial perturbations into the reference audio. 
Specifically, to obfuscate speaker identity, VoiceCloak first targets speaker identity by distorting representation learning embeddings to maximize identity variation, which is guided by auditory perception principles.
Additionally, VoiceCloak disrupts crucial conditional guidance processes, particularly attention context, thereby preventing the alignment of vocal characteristics that are essential for achieving convincing cloning. 
Then, to address the second objective, VoiceCloak introduces score magnitude amplification to actively steer the reverse trajectory away from the generation of high-quality speech. 
Noise-guided semantic corruption is further employed to disrupt structural speech semantics captured by DMs, degrading output quality. 
Extensive experiments highlight VoiceCloak's outstanding defense success rate against unauthorized diffusion-based voice cloning. 
Audio samples of VoiceCloak are available at \url{https://voice-cloak.github.io/VoiceCloak/}.


\end{abstract}


\section{Introduction}
\begin{figure}[!t] 
  \centering
  \includegraphics[width=0.97\linewidth]{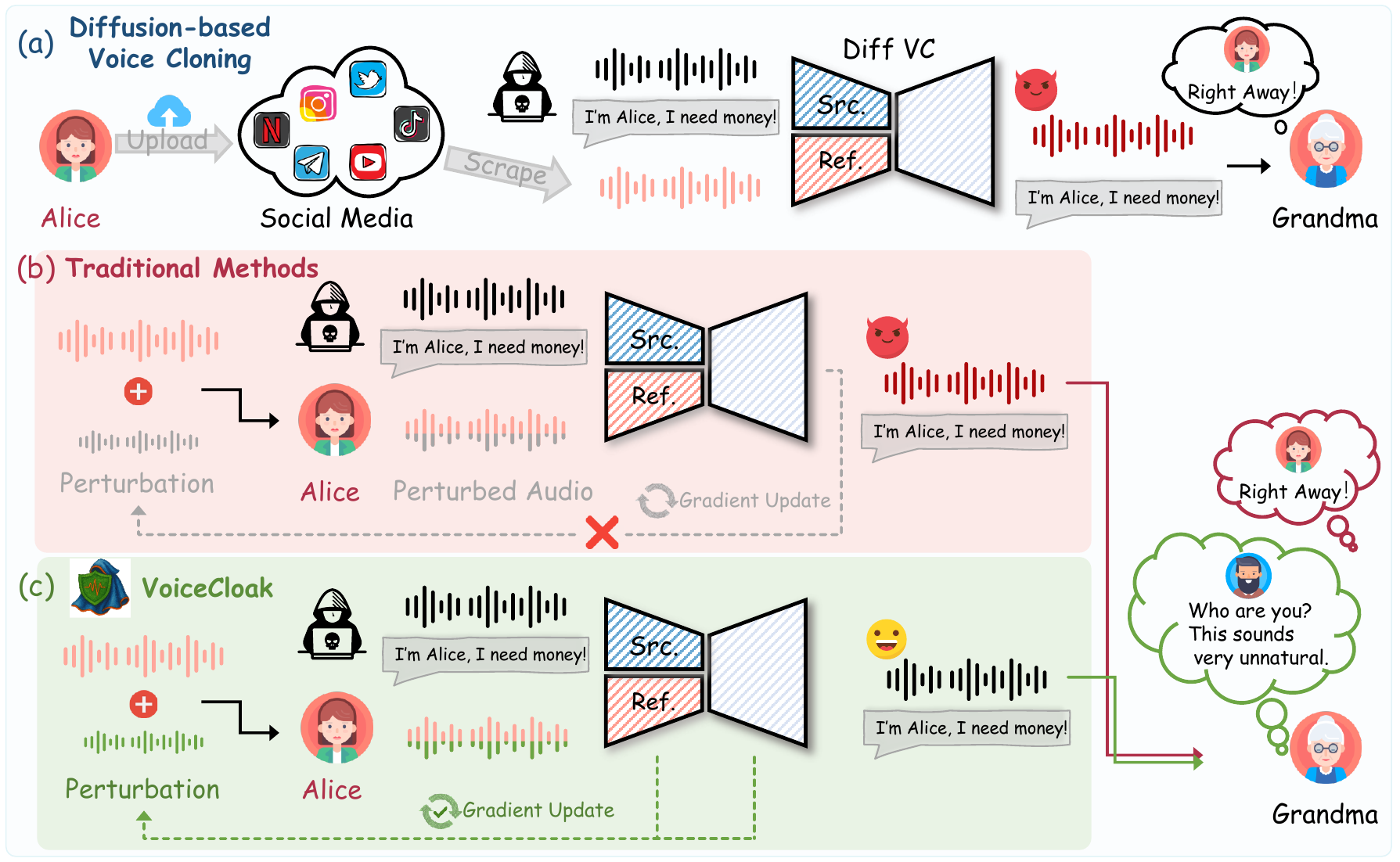}
  \caption{Illustration of diffusion-based voice cloning malicious misuse. (a) Voice forgery enables threats of fraud. (b) Traditional methods struggle due to ineffective disruptive gradients. (c) Audio protected by VoiceCloak resists high-fidelity cloning.}
  \label{fig:intro} 
\end{figure}

Diffusion Models (DMs) \cite{DDPM,DDIM,LDM} have recently emerged as powerful generative tools, achieving unprecedented success within realistic voice cloning (VC).
Their iterative denoising process enables generating speech with remarkable naturalness, detail, and fidelity to human voice \cite{DiffVC,DiffWave,DiffTTS,NaturalSpeech}. 
However, the open-source availability and ease of use of these models intensify concerns about potential misuse. Attackers can synthesize highly realistic voice replicas from short public audio clips, as depicted in Figure \ref{fig:intro}(a), enabling sophisticated fraud and circumvention of voiceprint authentication.

To counter such unauthorized use, two main defense paradigms are introduced, including forgery detection and proactive disruption. Reactive detection methods \cite{detection_1,detection_2,detection_3} identify forgeries after they are crafted, often too late to prevent harm. This highlights the need for proactive defenses that disrupt the synthesis process itself. 
Prior proactive work \cite{Attack-vc,AntiFake,voiceprivacy,VoiceGuard} has focused on adding imperceptible adversarial perturbations to reference audio by compromising the functionality of either the voice decoder or the speaker identity encoder.
However, existing defenses designed for prior architectures are largely ineffective against Diffusion Models (DMs). This incompatibility arises from two fundamental challenges: gradient vanishing and dynamic conditioning (Figure \ref{fig:intro}).
Specifically, (1) the single forward pass gradient computation relied upon by many defenses become unreliable or vanish within the multi-step denoising process of DMs and the corresponding deep computational graph, rendering such single-pass gradient information ineffective for disrupting the full generation trajectory \cite{Diffattack}.
(2) Strategies targeting specific subnetworks (e.g. speaker or content encoders) fail because DMs often employ dynamic conditioning mechanisms, which means no single modules solely responsible for condition processing. Consequently, methods targeting individual components struggle to cause global disruption.
These fundamental incompatibilities underscore the need for novel strategies tailored to this generative paradigm.

In light of the incompatibility of prior defenses with the diffusion paradigm, we introduce \textit{VoiceCloak}, a novel proactive defense framework designed for two primary objectives against unauthorized voice cloning: \textbf{Speaker Identity Obfuscation} and \textbf{Perceptual Fidelity Degradation}. Driven by these two objectives, we conduct an analysis to identify and exploit corresponding intrinsic vulnerabilities within Diffusion Models (DMs). Based on this analysis, we design specific optimization objectives for the protective perturbation to effectively disrupt the synthesis process.

To achieve identity obfuscation, VoiceCloak first directly manipulates speaker representations within a universal embedding space, guided by psychoacoustic principles to maximize perceived identity distance and hinder the DM's identity signature extraction. 
Second, recognizing that convincing mimicry depends on attention mechanisms in conditional guidance to align speaker style with content, we exploit this by introducing attention context divergence. This design prevents the attention mechanism from correctly utilizing contextual information, thereby disrupting the alignment required for accurate cloning.

Simultaneously, to achieve fidelity degradation, we focus on vulnerabilities within the core generative process itself. First, we employ Score Magnitude Amplification (SMA) to exploit the sensitivity of the iterative denoising trajectory which is crucial for realistic output, steering the generation path away from high-fidelity regions. 
Furthermore, acknowledging that the U-Net's understanding of high-level semantics governs output naturalness, we utilize noise-guided semantic corruption to disrupt the capture of structural features to promote incoherence within the noise semantic space and degrade generation quality. 
These goal-driven strategies which originates from adversarial analysis form a comprehensive defense. Extensive experiments confirm the superior defense efficacy against diffusion-based VC attacks under equivalent perturbation budgets.

Our contributions are summarized as follows:
\begin{itemize}
    \item We propose \textit{VoiceCloak}, a novel defense framework against Diffusion-Based VC that prevents unauthorized voice "theft" by exploiting intrinsic diffusion vulnerabilities through multi-dimensional adversarial interventions.
    \item We introduce auditory-perception-guided adversarial perturbations into speaker identity representations and disrupt the diffusion conditional guidance process to effectively distort identity information in synthesized audio.
    \item We present SMA, which controls the score function to divert the denoising trajectory, complemented by a semantic function designed to adversarially corrupt structural semantic features within the U-Net, degrading the perceptual fidelity of the forged audio.
\end{itemize} 

\section{Related Work}
\subsection{Audio Diffusion Models}
Diffusion Models \cite{related_DMs_1,related_DMs_2,related_DMs_3} a dominant force in generative modeling, demonstrating extraordinary performance across
 multi-modal tasks, significantly advancing audio synthesis tasks like text-to-speech \cite{GradTTS,DiffTTS} and unconditional audio generation \cite{DiffWave,AudioLDM}. Particularly within voice cloning (VC), diffusion-based methods\cite{DiffVC,dddm-vc}, mostly leveraging score-based formulations via stochastic differential equations (SDEs) \cite{SDE}, now yield outputs with remarkable naturalness and speaker fidelity. While impressive, this state-of-the-art performance significantly heightens concerns regarding potential misuse, directly motivating proactive defense strategies such as the framework proposed herein.
\subsection{Proactive Defense via Adversarial Examples}
Beyond passive DeepFake detection, proactive defenses aim to preemptively disrupt malicious syntheses, by introducing adversarial perturbations to the original audio.
Early work \cite{Attack-vc} demonstrated the feasibility of this approach but struggled to balance effectiveness with imperceptibility.
Subsequent research focused on improving this trade-off. Strategies included using psychoacoustic models to enhance imperceptibility \cite{VoiceGuard}, incorporating human-in-the-loop refinement for better balance \cite{AntiFake}, , and employing GAN-based generators to improve efficiency \cite{SWCSM}.


Despite these advancements, prior proactive strategies were largely designed for earlier generative architectures like GANs. They often overlook the unique mechanisms and internal structures of Diffusion Models (DMs), limiting their applicability. Recognizing this critical gap, our work proposes a defense specifically tailored to the challenges of diffusion-based voice cloning.


\section{Preliminaries}
\subsection{Score-based Diffusion} \label{sec:preliminary}
Score-based generative models define a continuous-time diffusion process using stochastic differential equations (SDEs) \cite{SDE}. The forward process gradually perturbs clean data $\mathbf{x}_0 \sim p_0(\mathbf{x})$ into noise via the SDE:
\begin{equation}
    d\mathbf{x} = f(\mathbf{x}, t) dt + g(t) d\mathbf{w},
    \label{eq:forward_diff}
\end{equation}
where $f(\mathbf{x},t)$ and $g(t)$ are the drift and diffusion coefficients, and $\mathbf{w}$ is a Wiener process. 
The corresponding \textbf{reverse-time SDE} that transforms $x_T$ back into $p_0(\mathbf{x})$ can be expressed as:
\begin{equation}
    d\mathbf{x} = \left[ f(\mathbf{x}, t) - g(t)^2 \nabla_{\mathbf{x}} \log p_t(\mathbf{x}) \right] dt + g(t) d\bar{\mathbf{w}},
    \label{eq:reverse_diff}
\end{equation}
where $\bar{\mathbf{w}}$ is the reverse-time Wiener process and $\nabla_{\mathbf{x}} \log p_t(\mathbf{x})$ represents the score-function. Then, the score-based diffusion model is trained to estimate the score function:
\begin{equation}
    \arg\min_{\theta} \lambda_t \mathbb{E}_{p_t(x)} \left\| s_\theta(\mathbf{x}_t, t) - \nabla \log p_t(\mathbf{x}_t|\mathbf{x}_0) \right\|^2_2,
    \label{eq:unet_train}
\end{equation}
where the expectation is taken over the data distribution $p_0(\mathbf{x}_0)$ and the transition kernel $p_t(\mathbf{x}_t|\mathbf{x}_0)$.

\subsection{Adversarial Vulnerability Analysis} \label{sec:analysis}
As mentioned before, the optimization of perturbation is guided by two core objectives: (1) Speaker Identity Obfuscation and (2) Perceptual Fidelity Degradation. The design of these perturbations stems from a targeted analysis to identify and exploit specific vulnerabilities within the Diffusion Model (DM) generative process itself. 

A primary objective of VoiceCloak is speaker identity obfuscation. Convincing identity mimicry in Diffusion Models (DMs) depends on precise conditional control which is guided by acoustic details modeled from the reference audio $x_{ref}$. This guidance critically relies on mechanisms that align the target speaker's acoustic characteristics from $x_{ref}$ with the phonetic content from $x_{src}$. Specifically, the attention block is responsible for executing this alignment. We identify this crucial acoustic style-to-content mapping as a key vulnerability,as inaccurate alignment directly compromises the successful rendering of the target speaker's identity. Consequently, disrupting this attention-driven pathway offers a strategy for identity obfuscation.

Complementary to identity obfuscation, the second objective is perceptual fidelity degradation, aimed at diminishing the usability of any synthesized speech. This involves targeting the core denoising process of DMs. The high fidelity of Diffusion-based VC relies on the model learning a precise reverse denoising trajectory to progressively refine noisy sample $x_t$ towards the natural audio distribution. This reliance on trajectory precision presents an exploitable vulnerability. Therefore, adversarially diverting the trajectory can disrupt convergence towards the high quality audio region on the manifold.

Additionally, we target the U-Net's internal feature representations as a vulnerability for degrading perceptual quality. Prior attribute editing work \cite{intro-UNet2, Attribute-edit1, Attribute-edit2} confirms that these hierarchical features within the U-Net are controllable and also encode crucial semantic and acoustic details that govern the coherence and naturalness of the synthesized speech. Therefore, adversarially corrupting these representations directly impair the model's ability to synthesize natural-sounding speech, offering a complementary strategy for fidelity degradation.
This analysis reveals critical vulnerabilities in DMs, providing targeted avenues for adversarial intervention.

\section{Methodology}
\begin{figure*}[!t] 
  \centering
  \includegraphics[width=0.95\linewidth]{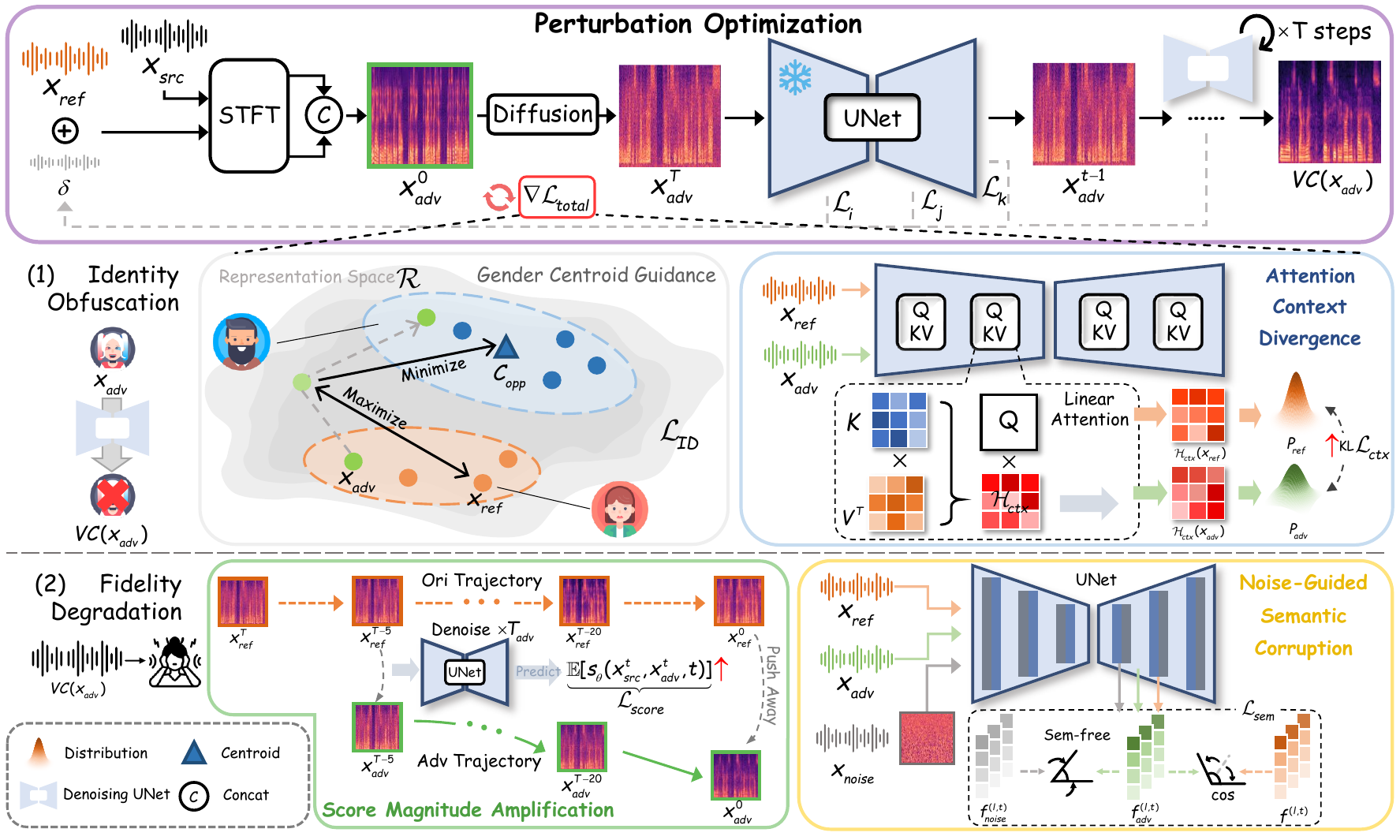}
  \caption{Overview of the proposed framework. Perturbation optimization is guided by gradients from $\mathcal{L}_{total}$, aggregating four targeting two objectives: (1) Identity Obfuscation (via Opposite-Gender Centroid Guidance and Attention Context Divergence) and (2) Perceptual Fidelity Degradation (Score Magnitude Amplification and Noise-Guided Semantic Corruption).}
  \label{fig:method_overview} 
\end{figure*}
To provide the necessary background, the problem formulation is firstly established in Section \ref{sec:problem_formula}. As shown in Figure \ref{fig:method_overview}, the VoiceCloak consists of four sub-modules, which will be introduced separately next. 
\subsection{Problem Formulation} \label{sec:problem_formula}
Assume that malicious users can obtain reference audio $x_{ref}$ of a target speaker. Leveraging open-source voice cloning models, they can synthesize speech which mimics the vocal characteristics of $x_{ref}$, denoted $\mathcal{VC}(x_{src},x_{ref},t)$. 
Our goal is to proactively safeguard $x_{ref}$ against unauthorized voice cloning. We achieve this by introducing an imperceptible adversarial perturbation $\delta$ to create a protected version $x_{adv}=x_{ref}+\delta$. This perturbation is optimized to disrupt the diffusion synthesis process when conditioned on $x_{adv}$. Formally, we aim to find an optimal $\delta$ that maximizes the dissimilarity between the outputs generated using $x_{ref}$ and $x_{adv}$.
\begin{equation}
    \begin{aligned}
         & \arg \max_{\delta} \  \mathcal{D}( \mathcal{VC}(x_{src},x_{ref},t),\ \mathcal{VC}(x_{src},x_{adv}, t) ), \\
        & \text{subject to} \quad  \Vert \delta \Vert_{\infty} \leq \epsilon,
    \end{aligned}
    \label{eq:formulation}
\end{equation}
where $\mathcal{D}(\cdot)$ measures the output discrepancy, $t$ is the diffusion timestep, and $\epsilon$ is the $l_{\infty}$-norm budget for the perturbation $\delta$. Our approach focuses on designing specific adversarial objectives that implicitly define $\mathcal{D}$ by exploiting intrinsic vulnerabilities within the diffusion mechanism itself.

\subsection{Adversarial Identity Obfuscation} \label{sec:identity}
\subsubsection{Opposite-Gender Embedding Centroid Guidance}
Inspired by prior work showing dedicated speaker embeddings capture identity \cite{voiceprivacy}, directly manipulating these embeddings offers a direct approach, but its practical utility is hindered by poor transferability when attacking unknown or different encoder models.

Therefore, we explore leveraging speech Representation Learning \cite{ReprLearning} to extract general acoustic representations that inherently capture speaker identity cues.
Specifically, we select WavLM \cite{Wavlm} as our representation extractor, denoted as $\mathcal{R}(\cdot)$. By applying perturbations within this more general feature space, we aim for broader effectiveness against various models. As a baseline untargeted objective, we consider maximizing the embedding distance:
\begin{equation}
    \mathcal{L}_{ID}=1 - Sim(\mathcal{R}(x_{adv}),\mathcal{R}(x_{ref})),
    \label{eq:untar_representation}
\end{equation}
where $Sim(x_1,x_2)$ represents the cosine similarity metric. 
While this untargeted objective effectively pushes $x_{adv}$ away from the reference identity representation domain, it lacks specific directionality in high-dimensional space. So we further incorporate a targeted component guided by psychoacoustic principles.

Psychoacoustic studies \cite{auditory-percep, F0_1} indicate that significant perceptual differences in speaker identity often exist between genders, linked to specific acoustic cues like F0 and formant structures . 
Leveraging this, we assume that guiding the adversarial embedding towards the opposite gender one will likely create the strongest perceptual contrast, enhancing identity obfuscation.

Based on this insight, we propose an auditory-perception-guided adversarial perturbation. Specifically, we first randomly select a speaker from the dataset whose gender is opposite to that of $x_{ref}$. 
To establish a representative identity embedding, we compute the centroid $\mathcal{C}_{opp}$ by averaging the embeddings of all utterances in $\mathcal{X}$ within the WavLM feature space $\mathcal{R}$:
\begin{equation}
    \mathcal{C}_{opp} = \frac{1}{N} \sum_{x_i \in \mathcal{X}} \mathcal{R}(x_i),
\end{equation}
 \begin{equation}
    \mathcal{L}_{ID} = - Sim(\mathcal{R}_{adv},\mathcal{R}_{ref})+ \underbrace{Sim(\mathcal{R}_{adv},\mathcal{C}_{opp})}_{\text{Gender}},
    \label{eq:loss_id}
\end{equation}
where $\mathcal{X}$ is the set of opposite gender utterances, $N$ represents length of $\mathcal{X}$, $\mathcal{R}_{i}$ is the embedding of $x_i$ in space $\mathcal{R}$.

Finally, minimizing the Eq. \ref{eq:loss_id} directs the optimization process to simultaneously dissociate from the original speaker identity and converge towards a region selected based on psychoacoustic principles to maximize identity ambiguity.

\subsubsection{Attention Context Divergence} \label{subsec:attention}
Building upon the motivation to disrupt conditional guidance for identity obfuscation (Section \ref{sec:analysis}), we introduce Attention Context Divergence. This strategy targets the attention mechanism to interfere with its use of contextual information from $x_{ref}$.

Commonly in diffusion-based VC, conditional information from $x_{ref}$ is integrated via Linear-attention layers \cite{LinAttention}. 
Within the U-Net, latent code representing the content $x_{src}^{t}$ are linearly projected to $Q$ matrix, whereas condition latents $x_{ref}^{t}$ are projected to $K$ and $V$. Linear-attention first computes a context matrix by aggregating values ($V$) weighted by keys ($K$):
$$
    \mathcal{H}_{ctx}(x_{ref})=\text{Softmax}(\phi^{(l,t)}(x_{ref}^t) W_{K}^{l})(\phi^{(l,t)}(x_{ref}^t) W_{V}^{l})^T,
$$
where $\mathcal{H}_{ctx}$ represents the context hidden state, $\phi^{(l,t)}(\cdot)$ is the deep features of the $l^{th}$ block in U-Net at timestep $t$, and $W_{K}^{l}, W_{V}^{l}$ are the projection matrices. Then, the context interacts with queries ($Q$) to obtain the final attention output: $\mathcal{A}^l = (\mathcal{H}_{ctx})^T(W_{Q}^{l}\phi^{(l,t)}(x_{src}^t))$.
This context representation provides a dynamic summary of the reference speaker's stylistic features, weighted by their relevance to the current content queries during synthesis.
By applying a softmax function, we obtain an explicit probability distribution over sequence positions which signifies the model's information focus. Our strategy, therefore, is to maximize the Kullback-Leibler (KL) divergence \cite{KL} between the context distribution derived from the original reference and the adversarial audio:
\begin{equation}
    \begin{aligned}
        \mathcal{L}_{ctx} &= D_{KL}(P_{ref} \parallel P_{adv}),  \\
         \text{where} \ P_{adv} &= \text{Softmax}(\mathcal{H}_{ctx}(x_{adv})),
    \end{aligned}
    \label{eq:kl_loss}
\end{equation}
Maximizing this divergence forces the attention pattern to deviate from the original, thereby hindering accurate style alignment.

To enhance the impact on identity, we directs the adversarial pressure on the U-Net's downsampling path. The rationale is that these earlier layers primarily process coarser, lower-frequency features \cite{simac} strongly associated with speaker timbre and identity.
Therefore, focusing our proposed loss $\mathcal{L}_{ctx}$ on the attention layers within the U-Net's downsampling path, the calculation of $P_{adv}$ can be restated as:
\begin{equation}
     P_{adv} = \text{Softmax}( \sum_l \mathcal{H}_{ctx}^l(x_{adv}) ),
\end{equation}
where the layer index $l$ iterates over the set $\text{Down}$ ($l\in\text{Down}$), representing the U-Net's downsampling blocks.

\subsection{Perceptual Fidelity Degradation} \label{sec:fidelity}
\subsubsection{Score Magnitude Amplification} \label{subsec:score}


To degrade perceptual fidelity by exploiting the sensitivity of the denoising trajectory, we introduce Score Magnitude Amplification. This design directly interferes with the score function $s_{\theta}$, which is estimated by the U-Net according to Eq. \ref{eq:unet_train} and provides the essential drift term for the reverse SDE.
We posit that the magnitude of $s_{\theta}$ relates to the strength of the drift guiding the noisy sample toward the target data manifold. Exploiting this connection, the SMA objective involves maximizing the magnitude of the score prediction $s_{\theta}$:
\begin{equation}
    \mathcal{L}_{score}= \mathbb{E}_{p_t(x),t \sim \mathcal{U}(1,T_{adv})} [\Vert s_{\theta}(x_{src}^t,x_{adv}^t,t) \Vert_{2}],
\end{equation}
where $p_t(x)$ is the distribution of noisy samples $x^t \sim q(x^t|x^0)$, $\mathbb{E}[\cdot]$ calculates the average value, $T_{adv}$ stands for adversarial timesteps which will be discussed in Section \ref{sec:overall}. 
Optimizing the above formula introduces an erroneous drift strength. Consequently, the denoising trajectory is forcefully diverted, resulting in a collapse in perceptual quality.

Furthermore, the iterative nature of the diffusion process may amplify these induced trajectory deviations. Perturbations introduced at earlier timesteps can propagate through subsequent steps. This error cumulative effect thus enhances the efficacy of our adversarial strategy.

\subsubsection{Noise-Guided Semantic Corruption} \label{subsec:semantic}
Following the motivation outlined in Section \ref{sec:analysis}, we introduce a bidirectional semantic interference strategy. The core idea is twofold: (1) compel the features generated with $x_{adv}$ to diverge from those generated using the original reference $x_{ref}$, and (2) concurrently guide these adversarial features towards a "semantic-free" state. 

Specifically, consider a network layer $l$ within the frozen U-Net $\mathcal{U}_{\theta}$ and timestep $t$, we extract the original features $f^{(l,t)} = \mathcal{U}_{\theta}^{l}(x_{src},x_{ref},t)$ conditioned on $x_{ref}$, and $ f^{(l,t)}_{adv} = \mathcal{U}_{\theta}^{l}(x_{src},x_{adv},t)$ corresponding to the adversarial version. Furthermore, to define a "semantic-free" target, we leverage the U-Net's activation modes to unstructured information. We extract features $f^{(l,t)}_{noise} = \mathcal{U}_{\theta}^{l}(x_{noise},x_{noise},t)$ by feeding Gaussian white noise $x_{noise}$ as both the source content and the reference condition. $f^{(l,t)}_{noise}$ can be considered to represent unstructured features and lack semantic information.
The bidirectional objective aims to maximize the distance between $f^{(l,t)}_{adv}$ and $f^{(l,t)}$ while minimizing it with $f^{(l,t)}_{noise}$. This objective encourages the adversarial features to abandon the original semantic structure and move towards a state of incoherence which can be formalized as:
\begin{equation}
    \mathcal{L}_{sem} = 1 - \cos(f^{(l,t)}_{adv},f^{(l,t)})+\underbrace{\cos(f^{(l,t)}_{adv},f^{(l,t)}_{noise})}_{\text{Sem-free}},
    \label{eq:semantic_loss}
\end{equation}
where we employ the cosine distance metric $\cos(\cdot)$ as it emphasizes the structural similarity between high-dimensional features rather than their absolute error. For enhanced impact on perceptual quality, the $\mathcal{L}_{sem}$ is strategically applied to layers within the U-Net's upsampling path. These layers are critical for reconstructing the fine-grained acoustic details that govern output naturalness and perceived quality.



\subsection{Joint Optimization of Defense Objectives} \label{sec:overall}
The final adversarial perturbation $\delta$ for VoiceCloak is optimized to simultaneously achieve our dual objectives (Section \ref{sec:identity} and \ref{sec:fidelity}). The joint objectives of this comprehensive defense are formalized as follows
\begin{equation}
    \begin{aligned}
        \mathcal{L}_{total} =& (\mathcal{L}_{ID},\mathcal{L}_{ctx},\mathcal{L}_{score},\mathcal{L}_{sem}) \Lambda^{T}, \\
        \delta :=& \ \arg \max_{\delta} \  \mathcal{L}_{total},
    \end{aligned}
    \label{eq:total_loss}
\end{equation}
where $\Lambda=(\lambda_{ID},\lambda_{ctx},\lambda_{score},\lambda_{sem})$ controls the weight factors that balance the relative importance of these defenses.

The efficacy of our perturbation optimization is influenced by the choice of diffusion timesteps $T_{adv}$ used for gradient computation. Informed by prior work \cite{timestep} indicating early denoising steps primarily reconstruct low-frequency overall structural signal, we concentrate the optimization on these initial steps to maximize the disruption of fundamental integrity and reduce computational overhead.
{
    \setlength{\heavyrulewidth}{1.2pt} 
    \begin{table*}[!t] 
      \centering

        \resizebox{0.85\textwidth}{!}{
            \begin{tabular}{c |l| ccccc | ccc}
            \toprule
            \multirow{2}{*}{\textbf{Datasets}} & \multirow{2}{*}{\textbf{Methods}} & \multicolumn{5}{c|}{\textbf{Defense Effectiveness}} & \multicolumn{3}{c}{\textbf{Imperceptibility}} \\
            & & \textbf{DTW$\uparrow$} & \textbf{ASV$\downarrow$} & \textbf{SSIM$\downarrow$} & \textbf{NISQA$\downarrow$} & \textbf{DSR$\uparrow$} & \textbf{PESQ$\uparrow$} & \textbf{MCD$\downarrow$} & \textbf{SNR$\uparrow$} \\
            \midrule
        
            \multirow{6}{*}{LibriTTS}
            & Undefended         & -    & 76.49\%   & -    & 3.96 & -         & -    & -    & -     \\
            & Random Noise $l_\infty$ & 2.01 & 55.20\%   & 0.31 & 3.72 & 16.00\%   & \textbf{3.37} & \underline{1.35} & \textbf{34.80} \\
            & Attack-VC               & \textbf{2.29} & 36.20\%   & 0.31 & \underline{3.57} & 30.40\%   & 2.31 & 3.71 & 5.29  \\ 
            & VoicePrivacy            & \underline{2.26} & 20.80\%   & 0.30 & 3.60 & 26.80\%   & 2.99 & 1.37 & 33.25 \\
            & VoiceGuard              & 2.08 & \underline{16.49}\%   & \underline{0.29} & 3.63 & \underline{43.45}\%   & 2.15 & 4.39 & 10.58 \\
            & \cellcolor[rgb]{ .906,  .902,  .902}\textbf{Ours} 
            & \cellcolor[rgb]{ .906,  .902,  .902}2.12 
            & \cellcolor[rgb]{ .906,  .902,  .902}\textbf{11.40\%} 
            & \cellcolor[rgb]{ .906,  .902,  .902}\textbf{0.27} 
            & \cellcolor[rgb]{ .906,  .902,  .902}\textbf{2.36} 
            & \cellcolor[rgb]{ .906,  .902,  .902}\textbf{71.40\%} 
            & \cellcolor[rgb]{ .906,  .902,  .902}\underline{3.22} 
            & \cellcolor[rgb]{ .906,  .902,  .902}\textbf{1.29} 
            & \cellcolor[rgb]{ .906,  .902,  .902}\underline{33.53} 
            \\ 
            \midrule
        
            \multirow{6}{*}{VCTK}
            & Undefended           & -    & 63.68\%   & -    & 3.41 & -         & -    & -    & -     \\
            & Random Noise $l_\infty$ & 1.68 & 58.00\%   & 0.35 & 3.16 & 11.38\%   & \textbf{3.25} & \underline{1.38} & \textbf{34.10} \\
            & Attack-VC               & \textbf{2.05} & 38.50\%   & 0.33 & 2.82 & 26.20\%   & 2.25 & 3.80 & 4.50  \\ 
            & VoicePrivacy            & 1.88 & \underline{30.28\%}   & 0.35 & \underline{2.77} & \underline{39.06\%}  & 2.87 & 1.53 & 31.87 \\
            & VoiceGuard              & 1.87 & 31.42\%   & \underline{0.32} & 3.02 &  22.11\%  & 2.05 & 4.57 & 10.00 \\
            & \cellcolor[rgb]{ .906,  .902,  .902}\textbf{Ours} 
            & \cellcolor[rgb]{ .906,  .902,  .902}\underline{1.93} 
            & \cellcolor[rgb]{ .906,  .902,  .902}\textbf{19.74\%} 
            & \cellcolor[rgb]{ .906,  .902,  .902}\textbf{0.29} 
            & \cellcolor[rgb]{ .906,  .902,  .902}\textbf{2.51} 
            & \cellcolor[rgb]{ .906,  .902,  .902}\textbf{63.41\%} 
            & \cellcolor[rgb]{ .906,  .902,  .902}\underline{3.09} 
            & \cellcolor[rgb]{ .906,  .902,  .902}\textbf{1.33} 
            & \cellcolor[rgb]{ .906,  .902,  .902}\underline{32.41} 
            \\ 
            \bottomrule
            \end{tabular}
        } 
        \caption{Comparison of defense effectiveness and adversarial imperceptibility with SOTA methods. Higher values are better for metrics marked with $\uparrow$, and vice versa for those marked with $\downarrow$. The best result is marked in \textbf{BOLD}.}
      \label{tab:comparison_main} 
    \end{table*}
}

\section{Experiments}
\subsection{Experimental Setup}
\subsubsection{Datasets}
Experiments are conducted on the LibriTTS \cite{LibriTTS} and VCTK \cite{VCTK} datasets. We selected a gender-balanced audio subsets (479 utterances from LibriTTS, 500 from VCTK) to generate adversarial reference speech set $\mathcal{D}_{x_{adv}}$.


\subsubsection{Baseline Methods}
We compare VoiceCloak against existing voice protection methods. 
We adopt the following methods for fair comparison: Attack-VC \cite{Attack-vc}, VoicePrivacy \cite{voiceprivacy}, and VoiceGuard \cite{VoiceGuard}. We also include a naive baseline: adding random Gaussian noise to $x_{ref}$.

\subsubsection{Evaluation Metrics}
For identity protection, we report the Automatic Speaker Verification (ASV) acceptance rate, where a lower rate for protected outputs signifies more effective obfuscation.
We define a comprehensive Defense Success Rate (DSR) to measure the achievement of both our objectives. A defense is considered successful if the protected output both fails speaker verification and exhibits low perceptual quality ($\mathrm{DS}= s_{ASV} < \tau_{ASV} \land \mathrm{NISQA}(y_{adv}) < \tau_{q}$), with thresholds $\tau_{ASV}=0.25$ and $\tau_{q}=3.0$.
Additional metrics include Dynamic Time Warping (DTW) \cite{DTW} and SSIM between $y$ and $y_{adv}$ spectrograms. Perturbation imperceptibility on $x_{adv} $ versus $ x_{ref}$ is measured using PESQ \cite{PESQ}, Mel-Cepstral Distortion (MCD) \cite{MCD} and SNR.

\subsubsection{Implementation Details}
We conduct our experiments mainly using DiffVC \cite{DiffVC} as the target system.
We set the number of optimization iterations to 50 and a step size $\alpha=4 \times 10^{-5}$. And $\delta$ is constrained within an $l_{\infty}$-norm ball of $\epsilon=0.002$. We set the adversarial and inference timesteps respectively $T_{adv}=6$ and $T=100$. The loss function combined identity and quality objectives with weights $\Lambda=(1.0, 4.5,10, 0.85)$. All experiments are conducted on NVIDIA RTX 3090 GPU with a fixed random seed.
\subsection{Comparison and Analysis}
\subsubsection{Comparison with Baselines}

\begin{figure}[!t] 
  \centering
  \includegraphics[width=0.97\linewidth]{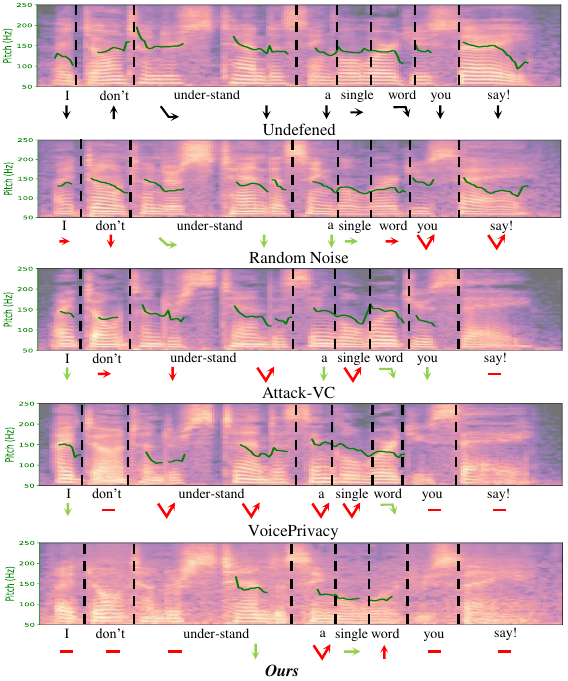}
  \caption{Mel spectrograms with $F_0$ pitch contours (\textcolor[rgb]{0.0, 0.4, 0.0}{green lines}), and inferred intonation of the corresponding words. Arrows indicate perceived intonation shifts. (Intonation \textcolor[rgb]{0.0, 0.4, 0.0}{aligns with} the ground truth, which is marked by \textcolor[rgb]{0.0, 0.4, 0.0}{green arrows}, and \textcolor{red}{diverges}, which is marked by \textcolor{red}{red arrows}.)}
  \label{fig:comparison_melspec} 
\end{figure}

As shown in Table \ref{tab:comparison_main}, VoiceCloak demonstrates exceptional defense efficacy, significantly outperforming all baselines with a Defense Success Rate (DSR) of over 71\% on LibriTTS and 63\% on VCTK. This high DSR reflects success in both of our defense objectives: identity obfuscation is achieved by drastically reducing speaker verification acceptance rates to ~11\%, while perceptual quality degradation is confirmed by low NISQA scores indicating unacceptable audio generation quality.
This dual ability to effectively cripple both identity mimicry and audio usability distinguishes VoiceCloak from defenses that may struggle with diffusion models or focus primarily on one objective.
As expected, the naive Random Noise baseline confirms that unstructured noise provides very limited protection.
Regarding imperceptibility, VoiceCloak performs comparably to baselines. This supports our strategy of targeting intrinsic vulnerabilities, rather than simply increasing perturbation magnitude.

Figure \ref{fig:comparison_melspec} further visualizes disruption result. Applying VoiceCloak protection results markedly different from the undefended one. The $F_0$ curve is notably degraded, appearing blurred and unpredictable, accompanied by highly inconsistent intonation changes.
\subsubsection{Protecting Commercial Systems}
\begin{figure}[!t] 
  \centering
  \includegraphics[width=0.96\linewidth]{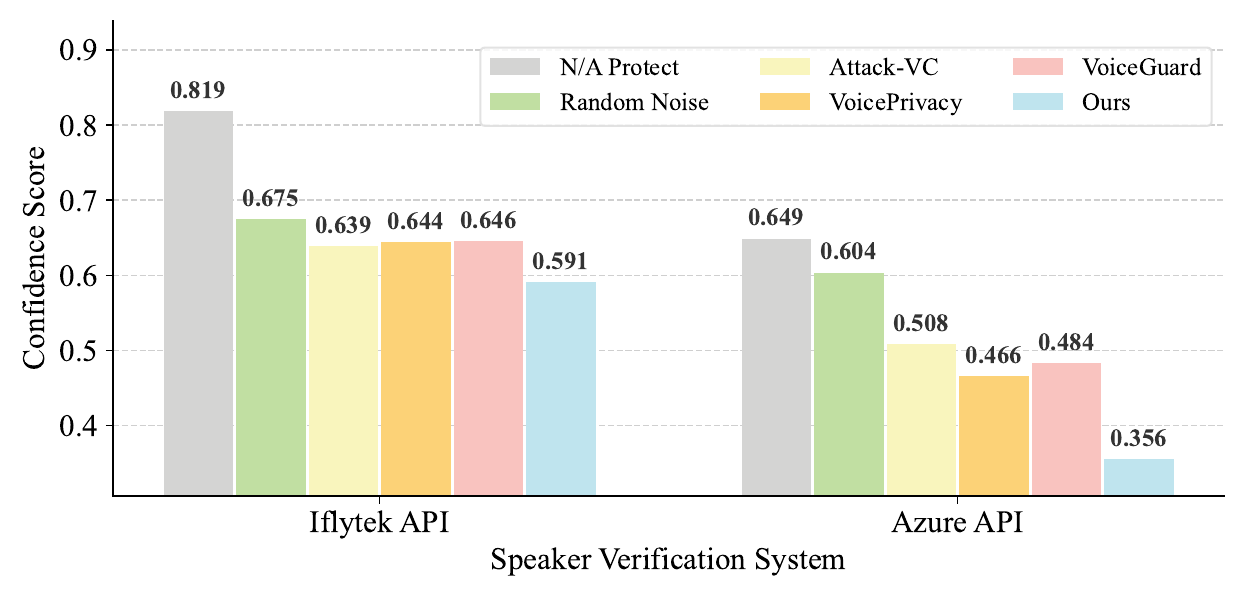}
  \caption{Protecting commercial speaker verification APIs (Iflytek, Azure) from spoofing attacks (lower are better).}
  \label{fig:commercial_APIs} 
\end{figure}

We also evaluated the effectiveness in protecting commercial speaker verification (SV) APIs (Iflytek, Azure) to simulate real-world anti-spoofing scenarios. Successful protection aims to minimize the similarity score returned by the API. Figure \ref{fig:commercial_APIs} demonstrated VoiceCloak's superior ability to decouple the protected audio from the original speaker's identity.

\subsubsection{User Study} \label{sec:user_study}

\begin{figure}[!t] 
  \centering
  \includegraphics[width=0.96\linewidth]{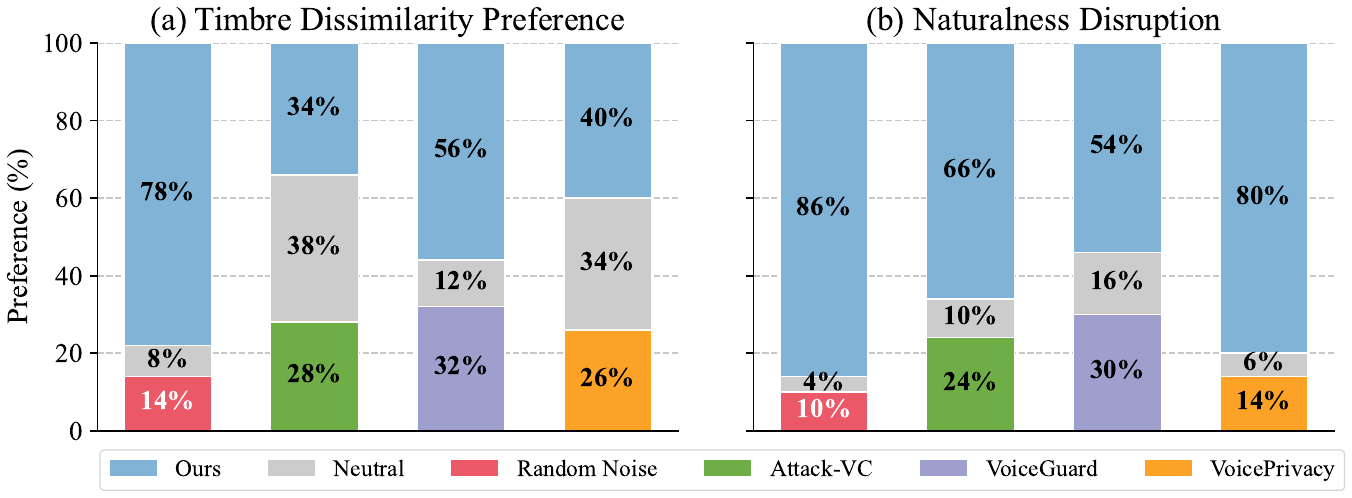}
  \caption{User perceptual study results. (a) Timbre Dissimilarity Preference. (b) Corresponding results for perceived Naturalness Disruption.}
  \label{fig:user_study} 
\end{figure}
We conducted a user study with 50 participants to assess perceptual impact. Listeners performed comparisons on two criteria: Timbre Dissimilarity and Naturalness Disruption and we aggregated results in Figure \ref{fig:user_study}, where "Neutral" indicates no perceived difference. Participants consistently rated VoiceCloak's outputs as having both greater timbre dissimilarity and more severe naturalness disruption, confirming its human-perceived effectiveness.
{
    \setlength{\heavyrulewidth}{1.2pt}
    \begin{table}[!t] 
      \centering
    
        \resizebox{0.94\linewidth}{!}{
        \begin{tabular}{cc |ccccc}
        \toprule
        \multicolumn{2}{c|}{\textbf{Settings}} & \multicolumn{5}{c}{\textbf{Defense Effectiveness}}  \\
        \textbf{$\mathcal{L}_{ID}$} & \textbf{$\mathcal{L}_{ctx}$} & \textbf{DTW$\uparrow$} & \textbf{ASV$\downarrow$} & \textbf{SSIM$\downarrow$} & \textbf{NISQA$\downarrow$} & \textbf{DSR$\uparrow$}  \\
        \midrule
        -               & -          & 1.96 & 46.82\% & 0.30 & 3.66 & 22.58\% \\
        $\checkmark$    & -          & 2.16 & 8.57\%  & 0.30 & 3.57 & 27.74\%  \\
        w/o $\text{Gender}$ & -          & 2.25 & 19.92\% & 0.30 & 3.60 & 14.40\%  \\
        -               & $\checkmark$ & 2.31 & 16.20\% & 0.27 & 2.96 & 62.57\%  \\
        $\checkmark$    & $\checkmark$ & 2.13 & 11.00\% & 0.27 & 2.85 & 69.20\%  \\
        \bottomrule
        
        \end{tabular}
        } 
        \caption{Ablation study on the contribution of different settings for Identity Obfuscation. $\checkmark$ indicates the setting is used, "w/o" denotes the exclusion of the specified term.}
        \label{tab:ablation_identity} 
    \end{table}
}
\subsection{Ablation Study}
\subsubsection{Effectiveness of Adversarial Identity Obfuscation}
{
    \setlength{\heavyrulewidth}{1.2pt}
    \begin{table}[!t] 
      \centering

        \resizebox{0.94\linewidth}{!}{
        \begin{tabular}{cc | ccccc}
        \toprule
        \multicolumn{2}{c|}{\textbf{Settings}} & \multicolumn{5}{c}{\textbf{Defense Effectiveness}}  \\
        \textbf{$\mathcal{L}_{score}$} & \textbf{$\mathcal{L}_{sem}$} & \textbf{DTW$\uparrow$} & \textbf{ASV$\downarrow$} & \textbf{SSIM$\downarrow$} & \textbf{NISQA$\downarrow$} & \textbf{DSR$\uparrow$} \\
        \midrule
        -            & -           & 1.99 & 45.00\% & 0.31 & 3.09 & 20.20\% \\
        $\checkmark$ & -           & 2.42 & 31.80\% & 0.29 & 2.68 & 41.20\% \\
        -            & $\checkmark$  & 2.23 & 23.00\% & 0.27 & 2.44 & 60.60\% \\
        -            & w/o Sem-free   & 2.28 & 26.36\% & 0.29 & 3.30 & 26.80\%  \\
        $\checkmark$ & $\checkmark$  & 2.22 & 23.60\% & 0.27 & 2.10 & 57.80\% \\
        \bottomrule
        \end{tabular}
        } 
         \caption{Ablation study for Perceptual Fidelity Degradation. Checkmark ($\checkmark$) indicates the setting is used,"w/o" denotes the exclusion of the specified loss term.}
         \label{tab:ablation_fidelity} 
    \end{table}
}
We ablate the loss designed for Adversarial Identity Obfuscation, $\mathcal{L}_{ID}$ and $\mathcal{L}_{ctx}$, in Table \ref{tab:ablation_identity}.
The results show that $\mathcal{L}_{ID}$ alone effectively reduces the ASV acceptance rate. This confirms that directly manipulating the representation learning embedding space effectively disrupts recognizable identity features.
Removing the opposite-gender guidance from $\mathcal{L}_{ID}$ worsens ASV, confirming our psychoacoustically-motivated strategy provides effective direction for identity disruption.
 Separately, the context divergence loss also contributes significantly, lowering both ASV and NISQA by interfering with the attention mechanism's condition injection.
\subsubsection{Effectiveness of Perceptual Fidelity Degradation}
Table \ref{tab:ablation_fidelity} ablates the perceptual quality degradation losses.
The results show that $\mathcal{L}_{score}$ alone is effective, degrading quality (lower NISQA, higher DTW) by forcing the denoising trajectory away from high-fidelity regions.
The semantic corruption loss, $\mathcal{L}_{score}$, demonstrates an even stronger individual impact by directly corrupting internal U-Net features, thereby impairing the model's reconstruction of coherent, natural-sounding speech details.
Critically, the necessity of guiding semantic features towards an incoherent state is evident from the "w/o Sem-free" variant.
\subsection{Transferability}
\begin{table}[!t]
  \centering
  
  \resizebox{0.92\linewidth}{!}{%
    \begin{tabular}{c| c c c c c} 
    \toprule
      \multirow{2}{*}{\textbf{Target Models}} & \multicolumn{5}{c}{\textbf{Defense Effectiveness}} \\
                       & DTW$\uparrow$ & ASV$\downarrow$ & SSIM$\downarrow$ & NISQA$\downarrow$ & DSR$\uparrow$ \\
      \midrule
      
      DiffVC           & 2.12 & 11.40\% & 0.27 & 2.36 & 71.40\% \\
      DDDM-VC          & 1.67 & 16.80\% & 0.36 & 2.79 & 54.89\% \\
      DuTa-VC          & 2.41 & 13.77\% & 0.27 & 2.14 & 73.92\% \\
      \bottomrule
    \end{tabular}%
  } 
  \caption{Transferability of the proposed defense method against unseen target models.}
  \label{tab:transferability}
\end{table}
To demonstrate applicability, we extended our experiments to include two additional open-source diffusion-based VC models: DuTa-VC \cite{Duta-vc} and DDDM-VC \cite{dddm-vc}. 
As shown in Table \ref{tab:transferability}, VoiceCloak demonstrate favorable transferability to different models, achieving an average DSR of 66.7\%. We attribute this transferability to: Targeting Common Vulnerabilities, as our method exploits fundamental mechanisms (e.g. attention, score prediction) that are shared across diffusion VCs.
\subsection{Robustness}
\begin{figure}[!t] 
  \centering
  \includegraphics[width=0.97\linewidth]{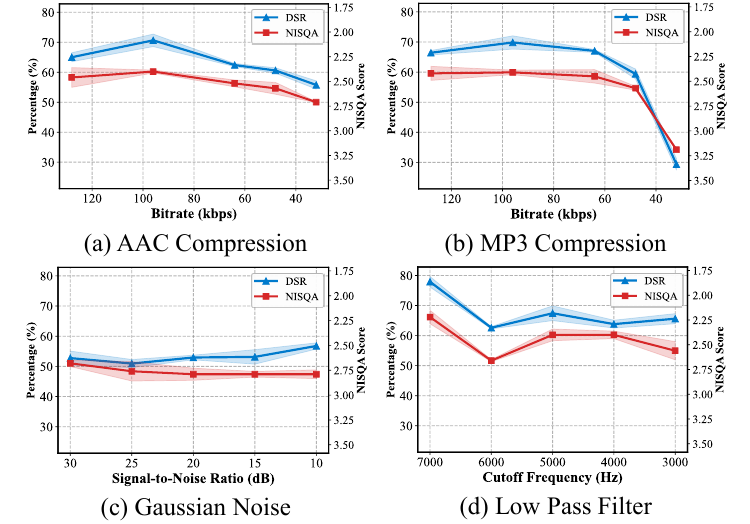}
  \caption{Resilience of VoiceCloak under more advanced robust scenario.}
  \label{fig: robustness} 
\end{figure}

We evaluated the robustness of our method against four common distortions, with results presented in Figure \ref{fig: robustness}. The results show consistent defense effectiveness, indicating resilience to real-world transformations and significantly outperforming the undefended baseline. Further details and analysis are available in Appendix.
\subsection{Efficiency Discussion}
\begin{table}[!t]
  \centering
  
  \resizebox{0.50\linewidth}{!}{%
      \begin{tabular}{c|c c}
      \toprule
            & \textbf{Time (s)} & \textbf{Mem. (GiB)} \\
        \midrule
        Ours  & 148.66            & 8.97                     \\
        \bottomrule
      \end{tabular}
  }
  \caption{Computational overhead analysis. "Time": average time to generate one sample. "Mem.": peak GPU usage.}
  \label{tab:computational_cost}
\end{table}
As shown in Table \ref{tab:computational_cost}, the protection process is a one-time, offline operation, and the required GPU memory is within the capacity of consumer-grade hardware, making VoiceCloak a practical defense method.
\section{Conclusion}

This paper introduced \textit{VoiceCloak}, a comprehensive defense against unauthorized diffusion-based voice cloning (VC). Our framework achieves superior defense effectiveness by exploiting targeted intrinsic vulnerabilities within the diffusion process. Through strategies designed to disrupt attention-based conditional guidance, steer the denoising trajectory, and corrupt internal semantic representations, \textit{VoiceCloak} effectively undermines the synthesis process. Extensive experiments validate its efficacy, demonstrating success in simultaneously obfuscating speaker identity and degrading audio quality to mitigate the threats of voice mimicry
Extensive experiments validate its ability to significantly hinder voice cloning by simultaneously disrupting identity and degrading audio quality, thereby mitigating the threats of high-quality voice mimicry

\putbib[aaai2026]
\end{bibunit}
\clearpage

\begin{bibunit}[aaai2026]  

\appendix
\section*{Technical Appendix}

\section{Additional Explanation on Our Strategy}
Algorithm. The full procedure of our multi-dimensional defense is detailed in Algorithm \ref{alg:voicecloak_final_io}
\begin{algorithm}[H]
    \caption{Adversarial Optimization Process on Diffusion-based VC}
    \label{alg:voicecloak_final_io}
    \begin{algorithmic}[1]
        \Statex \textbf{Input:} 
            Original reference audio $\mathbf{x}_{ref}$, source mel-spectrogram $\mathbf{m}_{src}$, opposite-gender centroid $\mathbf{C}_{opp}$, adversarial timestep $T_{adv}$, num of iterations $N$.
        \Statex \textbf{Result:} 
            Protected audio $\mathbf{x}_{adv}$.

        \State \textbf{Initialization:} $\mathbf{x}_{adv} \gets \mathbf{x}_{ref}$ 
        
        \State // \textit{Pre-compute features from original and noise inputs}
        \State $\mathbf{e}_{ref} \gets \mathcal{R}(\mathbf{x}_{ref})$
        \State $\mathcal{H}_{ctx}(x_{ref})=Attention(\mathbf{m}_{src}, \mathbf{x}_{ref}, T_{adv})$
        \State $f_{ref} \gets \mathcal{U}_{\theta}(\mathbf{m}_{src}, \mathbf{x}_{ref}, T_{adv})$ \Comment{Extract U-Net features}
        \State $\mathbf{x}_{noise} \gets \mathcal{N}(\mathbf{0}, \mathbf{I})$
        \State $\mathbf{f}_{noise} \gets \mathcal{U}_{\theta}(\mathbf{x}_{noise}, \mathbf{x}_{noise}, T_{adv})$

        \For{$n = 1 \dots N$}
            \State $\mathbf{e}_{adv} \gets \mathcal{R}(\mathbf{x}_{adv})$
            \State $\mathcal{H}_{ctx}(\textbf{x}_{adv}), \mathbf{f}_{adv} \gets \mathcal{VC}_(\mathbf{m}_{src}, \mathbf{x}_{adv}, T_{adv})$ \Comment{Extract all necessary features}
            \State $score \gets s_{\theta}(\textbf{x}_{ref},\textbf{x}_{adv},T_{adv})$
            \Statex // \textit{Compute loss components}
            \State $\mathcal{L}_{ID} \gets 1 - \text{SIM}(\mathbf{e}_{adv}, \mathbf{e}_{ref}) + \text{SIM}(\mathbf{e}_{adv}, \mathbf{C}_{opp})$
            
            \State $P_{ref} \gets \text{Softmax}(\mathcal{H}_{ctx}(\textbf{x}_{ref}))$
            \State $P_{adv} \gets \text{Softmax}(\mathcal{H}_{ctx}(\textbf{x}_{adv}))$
            \State $\mathcal{L}_{ctx} \gets D_{KL}(P_{ref} \parallel P_{adv})$
            
            \State $\mathcal{L}_{score} \gets ||score||_2$
            
            \State $\mathcal{L}_{sem} \gets (1 - \text{cos}(\mathbf{f}_{adv}, \mathbf{f}_{ref})) + \text{cos}(\mathbf{f}_{adv}, \mathbf{f}_{noise})$ 
            \Statex  // \textit{ Aggregate total loss and update via PGD}
            \State $\mathcal{L}_{total} \gets \lambda_{ID}\mathcal{L}_{ID} + \lambda_{ctx}\mathcal{L}_{ctx} + \lambda_{score}\mathcal{L}_{score} + \lambda_{sem}\mathcal{L}_{sem}$
            \State $\mathbf{x}_{adv} \gets \mathbf{x}_{adv} + \alpha \cdot \text{sign}(\mathbf{ \nabla_{\mathbf{x}_{adv}} \mathcal{L}_{total}})$
            \State $\mathbf{x}_{adv} \gets \mathbf{x}_{ref} + \text{clamp}(\mathbf{x}_{adv} - \mathbf{x}_{ref}, -\epsilon, \epsilon)$
        \EndFor
        
        \State \Return $\mathbf{x}_{adv}$
    \end{algorithmic}
\end{algorithm}
\section{Experimental Details}
\subsection{Baseline Methods}
To ensure a fair comparison, all baseline methods were configured in a white-box setting. For methods with official implementations (e.g., Attack-VC, VoiceGuard), we adapted their public code for our experimental pipeline. Perturbation budgets and key parameters for each baseline were carefully set, adhering to their original papers where possible.
\begin{itemize}
    \item Random Noise: A naive baseline where Gaussian noise ($ x \sim \mathcal{N}(0,I)$) is added to the reference audio. The noise is clipped to the same $l_{\infty}$-norm budget as our method for a direct comparison of structured vs. unstructured noise.
    \item Attack-VC \cite{attack-vc}: Following their paper, we combined two of their primary strategies: End-to-End (e2e): This defense directly maximizes the Mel-spectrogram $l_2$ distance between the VC outputs generated with the original and perturbed references which backpropagated through the entire Diffusion model. Speaker Embedding: This strategy targets a pre-trained d-vector as speaker encoder.
    \item VoicePrivacy \cite{voiceprivacy}: For this baseline, we strictly follow the original implementation from the original paper. Notably, this setup differs from the primary configuration used for our main experiments. Specifically, adversarial perturbations are generated using the Fast Gradient Sign Method (FGSM) with a prescribed budget of $\epsilon = 0.02$. The targeted speaker encoder, as specified in their work, is the pre-trained ECAPA-TDNN model provided with YourTTS. And the optimization objective is to maximize the corresponding cosine similarity between extracted embeddings. 
    \item VoiceGuard \cite{VoiceGuard}: VoiceGuard optimizes time-domain perturbations and uniquely constrains their magnitude using a psychoacoustic model rather than a fixed $l_p$-norm budget. Given this distinct constraint mechanism, a direct budget comparison is not applicable. Therefore, for a fair evaluation consistent with its design, we employed the default hyperparameter settings provided in the official codebase.
\end{itemize}

\subsection{Evaluation Metrics}
To provide a comprehensive and multi-faceted evaluation of VoiceCloak, we employ a suite of objective metrics assessing three key aspects: identity obfuscation, perceptual quality degradation, and perturbation imperceptibility. Let $y$ denote the audio synthesized using the original reference $x_{ref}$, and $y_{adv}$ denote the audio synthesized using the protected reference $x_{adv}$.

\textbf{Automatic Speaker Verification (ASV)}: This metric quantifies the core goal of identity obfuscation. We utilize a pre-trained, state-of-the-art ECAPA-TDNN \cite{ecapa} model as our speaker verification system. For each target speaker, we first compute an average enrollment embedding from set of their clean, original utterances. Then, for each defended output $y_{adv}$, we extract its embedding and compute a cosine similarity score against the corresponding target speaker's enrollment embedding. The ASV Acceptance Rate is samples whose similarity score exceeds a pre-defined threshold, which can be formulated as: $Sim(y_{adv},avg(X_{ref})) \geq \tau_{ASV}$.

\textbf{NISQA}: NISQA \cite{NISQA} is a neural network-based model for objective audio quality assessment. It is trained to predict the subjective Mean Opinion Score (MOS) that a human listener would assign, evaluating the overall quality and naturalness of speech on a scale of 1 to 5. By leveraging a deep model trained on human perception data, NISQA provides a robust and objective proxy for subjective listening tests.

\textbf{Dynamic Time Warping}: DTW \cite{DTW} measures the distance between the temporal structures of two audio signals. Based on our implementation, we first extract the Mel spectrograms of the undefended $y$ and defended $y_{adv}$ audio outputs. We then compute the \textit{cityblock} distance between these two spectrograms using the fastdtw algorithm to find the optimal alignment path. 

\textbf{Structural Similarity (SSIM)}: Originally an image quality metric, we adapt SSIM \cite{SSIM} to measure the perceptual similarity between the spectrograms of $y$ and $y_{adv}$. We first compute the Short-Time Fourier Transform (STFT) of both audio signals and convert the resulting magnitudes to a log-dB scale. These log-magnitude spectrograms are then normalized to a range of [0, 1]. The SSIM score is calculated between these normalized spectrograms. The SSIM index between two spectrograms is calculated as: 
\begin{equation}
    \text{SSIM}(u, v) = \frac{(2\mu_u\mu_v + C_1)(2\sigma_{uv} + C_2)}{(\mu_u^2 + \mu_v^2 + C_1)(\sigma_u^2 + \sigma_v^2 + C_2)},
\end{equation}
where $\mu$ and $\sigma$ represent the mean and standard deviation, respectively, $\sigma_{uv}$ is the covariance, and $C_1, C_2$ are stabilizing constants. A lower SSIM score (closer to 0) indicates less structural similarity and thus a greater defensive effect.

\textbf{Defense Success Rate (DSR)}: DSR is our primary comprehensive metric, designed to holistically evaluate the achievement of our two defense objectives. A defense instance is deemed successful only if the resulting audio $y_{adv}$ simultaneously meets two conditions: (1) it fails the speaker verification test (i.e., ASV score $\leq \tau_{ASV}$), and (2) it exhibits low perceptual quality. Eventually, a successful defense is formally defined as: 
$$\text{Success} = (\text{ASV}(y_{adv}) < \tau_{ASV}) \land (\text{NISQA}(y_{adv}) < \tau_{q}),$$
where we set the quality threshold $\tau_{q} = 0.30$ and identity threshold $\tau_{ASV} = 0.25$. The thresholds were chosen to reflect reasonable boundaries for identity verification and quality assessment. And these choices align with prior work or related paper, providing a consistent standard and  quantifiable success defense.

The following metrics evaluate the quality of the adversarial reference audio $x_{adv}$ itself, by comparing it to the original $x_{ref}$.

\textbf{Perceptual Evaluation of Speech Quality (PESQ)}: PESQ \cite{PESQ} is a classic objective metric that predicts the subjective quality of speech. We use it to measure the degradation introduced by the perturbation. Scores range from -0.5 to 4.5, with higher scores indicating better quality (i.e., a more imperceptible perturbation).

\textbf{Mel-Cepstral Distortion (MCD)}: This calculates the Euclidean distance between the Mel-frequency cepstral coefficients (MFCCs) of $x_{ref}$ and $x_{adv}$. It is a standard metric for measuring spectral distortion in speech. A lower MCD value indicates less spectral distortion and thus a more imperceptible perturbation.

\textbf{Signal-to-Noise Ratio}: SNR measures the ratio of the power of the original signal $x_{ref}$ to the power of the perturbation noise $\delta$. It is calculated in decibels (dB) as:
\begin{equation}
    \text{SNR}_{\text{dB}} = 10 \log{10} \left( \frac{\sum_{n} x_{ref}(n)^2}{\sum_{n} \delta(n)^2} \right)
\end{equation}
where the sums are over all samples in the signal. A higher SNR value indicates a weaker perturbation relative to the signal.

\subsection{Implementation Details}
\subsubsection{Hyperparameter Setup}
\begin{table}[!ht]
 \setlength{\heavyrulewidth}{1.17pt} 
  \centering
  \resizebox{0.95\linewidth}{!}{
      \begin{tabular}{cccc} 
        \toprule
        \textbf{Norm} & \textbf{$\epsilon$} & \textbf{step size $\alpha$} & \textbf{optimization iterations}\\
        \midrule
        $\ell_{\infty} $ & 0.002 & $4 \times 10^{-5}$ &  50 \\
        \bottomrule
      \end{tabular}
  }
    \caption{Hyperparameters used for the PGD.}
    \label{tab:pgd_hyperparameter}
\end{table}
All audio data was uniformly resampled to 22050 Hz and normalized to the range [-1, 1].
To generate the set of protected audio samples $x_{adv}$, we selected a diverse and gender-balanced subset of reference utterances. Specifically, from LibriTTS, we used 479 utterances, and from VCTK, we used 500 utterances. In both cases, the selection ensures an approximate 50\% male and 50\% female speaker distribution. 
For the voice cloning synthesis process, all voice cloning evaluations were conducted as gender-matched conversions (i.e., male source to male target, female source to female target), as this setup typically yields the highest baseline voice conversion quality, to ensure a consistent and challenging evaluation scenario.
\begin{table}[!ht]
 \setlength{\heavyrulewidth}{1.17pt} 
  \centering
  \resizebox{0.95\linewidth}{!}{
      \begin{tabular}{ccccc} 
        \toprule
        \textbf{mel bins} & \textbf{hidden dims} & \textbf{denoise timesteps} & \textbf{Adv. timesteps} \\
        \midrule
        80 & 256 &  100 & 6 \\
        \bottomrule
      \end{tabular}
  }
    \caption{Hyperparameters used for diffusion process.}
    \label{tab:diffusion_hyperparameter}
\end{table}

In this paper, our adversarial perturbations are generated using the Projected Gradient Descent (PGD) algorithm. The optimization process is configured with the hyperparameters detailed in Table \ref{tab:pgd_hyperparameter}.
For the diffusion process itself, we set the configuration with the hyperparameters, as shown in Table \ref{tab:diffusion_hyperparameter}.
{
    \begin{table*}[!t]
      \centering
     \setlength{\heavyrulewidth}{1.17pt} 
      \resizebox{0.67\linewidth}{!}{%
        \begin{tabular}{c|c| cccc} 
          \toprule
          \multirow{2}{*}{\textbf{Lossy Type}}& \multirow{2}{*}{\textbf{Level}($\uparrow$)} &\multicolumn{4}{c}{\textbf{Defense Effectiveness}} \\
          \cmidrule(lr){3-6}
            &  & \textbf{DTW}($\downarrow$) & \textbf{ASV}($\downarrow$) & \textbf{NISQA}($\downarrow$) & \textbf{DSR}($\uparrow$) \\
          \hline
          
          \multirow{5}{*}{AAC Compression} 
                           & 128 & 2.33 & 16.70\% & 2.46 & 65.01\% \\
                           & 96  & 2.33 & 14.08\% & 2.40 & 70.65\% \\
                           & 64  & 2.31 & 15.49\% & 2.52 & 62.40\% \\
                           & 48  & 2.30 & 19.92\% & 2.57 & 60.59\% \\
                           & 32  & 2.32 & 14.08\% & 2.71 & 55.76\% \\
          \hline
          \multirow{5}{*}{MP3 Compression} 
                           & 128 & 2.33 & 17.71\% & 2.42 & 66.42\% \\
                           & 96  & 2.33 & 13.48\% & 2.41 & 69.84\% \\
                           & 64  & 2.33 & 15.29\% & 2.45 & 67.03\% \\
                           & 48  & 2.31 & 17.91\% & 2.57 & 59.38\% \\
                           & 32  & 2.34 & 15.29\% & 3.19 & 29.40\% \\
          \hline
          \multirow{5}{*}{Gaussian Noise}  
                           & 30  & 2.34 & 19.72\% & 2.68 & 52.74\% \\
                           & 25  & 2.35 & 18.51\% & 2.76 & 50.93\% \\
                           & 20  & 2.37 & 16.50\% & 2.79 & 52.94\% \\
                           & 15  & 2.39 & 12.88\% & 2.79 & 53.14\% \\
                           & 10  & 2.47 & 10.66\% & 2.79 & 56.76\% \\
          \hline
          \multirow{5}{*}{Lowpass}         
                           & 7000 & 2.32 & 13.88\% & 2.22 & 77.89\% \\
                           & 6000 & 2.40 & 17.30\% & 2.66 & 62.60\% \\
                           & 5000 & 2.39 & 21.13\% & 2.40 & 67.43\% \\
                           & 4000 & 2.39 & 22.33\% & 2.40 & 63.81\% \\
                           & 3000 & 2.44 & 17.91\% & 2.56 & 65.62\% \\
          \bottomrule
        \end{tabular}%
      } 
      \caption{Overall robustness of our defense method against various lossy post-processing operations. Lower values for ``Level'' indicate more severe operations.}
      \label{tab:lossy_robustness}
    \end{table*}
}
\subsubsection{Human Evaluation}
 We conducted a subjective human evaluation study to assess the real-world perceptual impact of VoiceCloak compared to baseline methods.
We recruited 50 participants for the study. A total of 100 distinct audio clips were used for evaluation, comprising samples synthesized using references protected by our method and the primary baseline methods. All participants were asked to use headphones in a quiet environment to ensure a consistent and high-quality listening experience. Participants were then asked to make a comparative judgment in response to the following two questions:
(1) "Which of the two synthesized voices sounds less similar to the original speaker's voice?" This question directly measures the perceived effectiveness of identity obfuscation.
(2) "Which of the two synthesized voices sounds less natural or more distorted?" This question evaluates the perceived degree of perceptual quality degradation.
For each question, participants could choose one of the two synthesized audio clips or select "Neutral" if they perceived no significant difference. To mitigate ordering bias, the positions of the two synthesized samples were randomly shuffled for each question and each participant.

\section{Additional Experiments}
\subsection{Robustness Evaluation}
We evaluated the robustness of our adversarial perturbations against common audio distortions encountered during digital transmission and storage. We subjected the protected reference audio $x_{adv}$ to four types of transformations with varying levels of severity:
\begin{itemize}
    \item \textbf{AAC and MP3 Compression}: These are widely used lossy audio codecs that reduce file size by discarding less perceptible acoustic information. We tested a range of bitrates from 128kbps (high quality) down to 32kbps (low quality).
    \item \textbf{Gaussian Noise}: This involves adding white Gaussian noise to the audio signal, a common way to simulate channel noise or environmental interference. We evaluated across Signal-to-Noise Ratios (SNR) from 30dB (low noise) to 10dB (high noise).
    \item \textbf{Low-pass Filtering}: This process removes high-frequency components from the signal, which can occur during resampling or transmission over band-limited channels. We tested cutoff frequencies from 7kHz down to 3kHz.
\end{itemize}
The defense performance under these conditions is detailed in Table \ref{tab:lossy_robustness}. Even under moderate degradation, the core defensive properties of the perturbation are well-preserved.

A \textbf{notable phenomenon} occurs under severe distortion (e.g., 32kbps MP3 compression), where defense metrics paradoxically appear to improve. We attribute this not to an enhanced perturbation, but to the degradation of the reference audio's fundamental signal integrity. At such high distortion levels, essential acoustic features of the speaker are damaged alongside the perturbation. Consequently, the voice cloning model itself is inherently compromised by the poor input quality, struggling to synthesize a coherent, identity-consistent output. In these extreme cases, the high DSR reflects not only our method's robustness but also the diffusion model's own limitation when presented with severely degraded inputs.

{

\begin{table}[!ht]
  \centering
  \setlength{\heavyrulewidth}{1.17pt} 
 
  \resizebox{0.99\linewidth}{!}{%
    \begin{tabular}{c|c|ccc} 
      \toprule
      \multirow{2}{*}{\textbf{Lossy Type}} & \multirow{2}{*}{\textbf{Methods}} & \multicolumn{3}{c}{\textbf{Defense Effectiveness}} \\
      \cmidrule(lr){3-5} 
      & & \textbf{ASV}($\downarrow$) & \textbf{NISQA}($\downarrow$) & \textbf{DSR}($\uparrow$) \\
      \midrule
      
      \multirow{3}{*}{AAC-32} 
        & Attack-VC    & 23.54\% & 3.30 & 25.35\% \\
        & VoicePrivacy & 22.74\% & 3.28 & 24.35\% \\
        & Ours         & \textbf{14.08}\% & \textbf{2.71} & \textbf{55.76}\% \\
      \midrule
      
      \multirow{3}{*}{MP3-32} 
        & Attack-VC    & 21.73\% & 3.60 & 12.07\% \\
        & VoicePrivacy & 19.52\% & 3.66 & 11.47\% \\
        & Ours         & \textbf{15.29}\% & \textbf{3.19 }& \textbf{29.40}\% \\
      \midrule
      
      \multirow{3}{*}{Gaussian-10} 
        & Attack-VC    & 16.50\% & \textbf{2.79} & 56.14\% \\
        & VoicePrivacy & 16.90\% & 2.77 & \textbf{57.14}\% \\
        & Ours         & \textbf{10.66}\% & \textbf{2.79} & 56.76\% \\
      \midrule
      
      \multirow{3}{*}{Lowpass-3k} 
        & Attack-VC    & 25.15\% & 2.87 & 43.69\% \\
        & VoicePrivacy & 23.94\% & 2.89 & 47.89\% \\
        & Ours         & \textbf{17.91}\% & \textbf{2.56} & \textbf{65.62}\% \\
      \bottomrule
    \end{tabular}%
  } 
   \caption{Comparison with prior defense methods under the most severe lossy post-processing conditions.}
  \label{tab:sota_robustness_comparison}
\end{table}
}
{
    \begin{table*}[!t]
    \centering
    \setlength{\heavyrulewidth}{1.17pt} 
    \resizebox{0.53\linewidth}{!}{%
        \begin{tabular}{c| c c} 
        \toprule
        \textbf{Methods} & \textbf{Time (seconds/audio)} & \textbf{Peak Memory (GB)} \\
        \hline
        
        Attack-VC    & 181.35   & 9.80  \\
        VoicePrivacy      & 16.12    & 0.61   \\
        VoiceGuard   & 248.91   & 14.36 \\
        Ours         & 148.66   & 9.19 \\
        \bottomrule
        \end{tabular}
    }
    \caption{Comparison of computational overhead for our method and baseline.}
    \label{tab:computational_overhead}
    \end{table*}
}
Furthermore, Table \ref{tab:sota_robustness_comparison} compares the robustness of VoiceCloak against baseline methods under the most severe distortion conditions. The results show that VoiceCloak consistently demonstrates superior defense effectiveness. Notably, VoiceCloak also achieves the \textbf{lowest ASV acceptance} rate in all scenarios, indicating that the identity obfuscation component of our perturbation is particularly resilient to these transformations.
This superior robustness is likely attributed to our \textbf{multi-dimensional defense strategy}. While simple post-processing might partially neutralize attacks targeting a single vulnerability (e.g., only the embedding space), VoiceCloak's approach of simultaneously corrupting conditional guidance, the denoising trajectory, and internal semantic features creates a more resilient and multifaceted defense that is harder to fully compromise.
\subsection{Computational Costs}
The practical utility of a proactive defense method is closely tied to its computational cost. For a tool like VoiceCloak to be accessible to individual users, it must be both time-efficient enough for practical use and memory-efficient enough to run on common consumer-grade hardware. In this section, we analyze the computational overhead of our method and compare it against the baselines.

\subsubsection{Analysis of the Computational Cost in VoiceCloak}

\textbf{Time Cost.}  In each of the 50 iterations, the primary operations are: (1) a forward pass through the diffusion model to compute the intermediate features and the score prediction $s_{\theta}$, which is necessary for all our loss components; and (2) a backward pass to compute the gradient of the total loss $\mathcal{L}_{total}$ with respect to the input audio $x_{adv}$. We set the adversarial optimization iterations to be 50 and the gradient repeats to be 5 for an average gradient. Therefore, we run $50 \times 5=250$ steps for one input optimization. As we select the early stage of diffusion steps, it would be time-efficient for backward propagation.

\textbf{Memory Cost.} The peak GPU memory usage during this process consists of three main components: (1) the model weights of the DiffVC model and any auxiliary models (like WavLM), which must be loaded into VRAM; (2) the computation graph and intermediate activations stored during the forward pass, which are necessary for gradient computation in the backward pass; and (3) a smaller amount of memory for the input audio tensors and the calculated gradients themselves. The dominant factor for memory is the size of the computation graph, which is directly related to the depth and complexity of the DiffVC U-Net.

\subsubsection{Evaluation on Costs}

Table \ref{tab:computational_overhead} presents a comparison of the average time required to process a single audio sample (approximately 7 seconds in length) for our method and the baselines on an NVIDIA RTX 3090 GPU. VoiceCloak requires approximately 148.7 seconds per sample. This is significantly faster than VoiceGuard (248.9s) and Attack-VC (181.4s). While methods like VoicePrivacy (16.1s) are faster, they often target simpler objectives and may not provide the same level of multi-faceted defense. Crucially, as this is a one-time, offline pre-processing step, we argue that this time cost is a reasonable and practical investment for users seeking to permanently protect their audio data before dissemination.

The peak GPU memory usage is another critical factor for accessibility. The memory cost of VoiceCloak is primarily determined by the storage of the DiffVC model weights and the computation graph required for backpropagation. A key design choice that significantly enhances our memory efficiency is the targeted computation over early denoising timesteps. Unlike a hypothetical, fully end-to-end attack that would need to backpropagate through the entire multi-step denoising trajectory—a process requiring prohibitively large memory to store dozens of U-Net computation graphs (at least $8 \times$ Nvidia A100 with 40GB memory). By concentrating gradient computations on only a few initial steps ($T_{adv}=6$), VoiceCloak only needs to store a much shallower computation graph, dramatically reducing the memory footprint.
{
    \setlength{\heavyrulewidth}{1.17pt} 
        \begin{table*}[!h]
          \centering
          \resizebox{0.85\linewidth}{!}{%
                \begin{tabular}{c | ccccc | ccc} 
              \toprule
              \multirow{2}{*}{\textbf{Budgets ($\epsilon$)}} & \multicolumn{5}{c|}{\textbf{Defense Effectiveness}} & \multicolumn{3}{c}{\textbf{Imperceptibility}} \\
                               & \textbf{DTW}($\uparrow$) & \textbf{ASV}($\downarrow$) & \textbf{SSIM}($\downarrow$) & \textbf{NISQA}($\downarrow$) & \textbf{DSR}($\uparrow$) & \textbf{PESQ}($\uparrow$) & \textbf{MCD}($\downarrow$) & \textbf{SNR}($\uparrow$) \\
              \midrule
              
              Undefended       & -       & 76.49\% & -      & 3.96 & -       & -      & -      & - \\
              0.0005           & 2.01    & 22.00\% & 0.29   & 2.86 & 41.20\% & 4.25   & 0.48   & 45.03 \\
              0.0010           & 2.05    & 16.60\% & 0.28   & 2.65 & 57.00\% & 3.81   & 0.79   & 39.26 \\
              0.0020           & 2.12    & 11.40\% & 0.27   & 2.36 & 71.20\% & 3.21   & 1.29   & 33.53 \\
              0.0050           & 2.23    & 6.20\%  & 0.25   & 1.95 & 84.00\% & 2.34   & 2.34   & 25.94 \\
              0.0100           & 2.32    & 1.20\%  & 0.24   & 1.75 & 90.80\% & 1.74   & 3.40   & 20.18 \\
              \bottomrule
        \end{tabular}%
        } 
        \caption{Performance of VoiceCloak under different perturbation budgets. "Undefended" denotes voice conversion without VoiceCloak.}
        \label{tab:ablation_epsilon} 
    \end{table*}
}  
As shown in Table \ref{tab:computational_overhead}, the peak memory usage of VoiceCloak is approximately 9.19 GB. This is lower than both Attack-VC (9.80 GB) and VoiceGuard (14.36 GB), which require larger memory footprints. 
A memory requirement of ~9.2 GB ensures that VoiceCloak is well within the capacity of most modern consumer-grade GPUs (e.g., those with 10GB, 12GB, or more VRAM), making it a highly accessible tool for individual users and non-specialists. This practical memory efficiency is a key advantage for real-world deployment.
 
\subsection{Additional Ablation Studies}
\subsubsection{Perturbation budgets}

We investigated the influence of the perturbation budget $\epsilon$ on both defense effectiveness and adversarial imperceptibility, with results summarized in Table \ref{tab:ablation_epsilon}. The findings reveal a clear trade-off: increasing e directly enhances defense effectiveness but conversely degrades audio imperceptibility. The excessive distortion compromises the usability of $x_{adv}$ for legitimate purposes. We determined $\epsilon = 0.0020$ to be a suitable default budget, providing strong protection while preserving acceptable adversarial quality.

\subsubsection{Robust to Inference Steps}
In a practical defense scenario, the defender cannot know the specific number of inference steps ($T$) an attacker will use for the voice cloning synthesis. To evaluate the robustness of VoiceCloak against this uncertainty, we conducted an ablation study where the attacker's total number of inference timesteps was varied.
For this experiment, the gradient-based loss components for perturbation were computed only within the early diffusion timesteps ($T_{adv}=6$). We then evaluated the effectiveness of this single, fixed perturbation when the diffusion model used it to synthesize audio over different inference path lengths, from a very short 
$T=6$ to $T=200$. The results are presented in Table \ref{tab:ablation_inference_steps}.
{
    \begin{table}[!h]
      \centering
      \setlength{\heavyrulewidth}{1.17pt} 
      \resizebox{0.99\linewidth}{!}{%
        \begin{tabular}{c | cccc | c} 
          \toprule
          \multirow{2}{*}{\textbf{Infer. Steps $T$}} & \multicolumn{4}{c|}{\textbf{Defense Effectiveness}} & \multirow{2}{*}{\textbf{Avg. RT (s)}} \\
          \cmidrule(lr){2-5} 
                           & \textbf{DTW}($\uparrow$) & \textbf{ASV}($\downarrow$) & \textbf{NISQA}($\downarrow$) & \textbf{DSR}($\uparrow$) & \\
          \midrule
          
          6                & 1.72 & 11.80\% & 1.83 & 84.20\% & 0.32 \\
          18               & 1.94 & 10.40\% & 2.08 & 81.80\% & 0.71 \\
          30               & 2.03 & 11.40\% & 2.13 & 78.80\% & 1.09 \\
          100              & 2.13 & 11.00\% & 2.37 & 71.80\% & 3.28 \\
          200              & 2.15 & 13.40\% & 2.37 & 69.20\% & 6.57 \\
          \bottomrule
        \end{tabular}%
      } 
      \caption{Ablation study on the effect of different diffusion inference steps $T$. "Avg. RT" indicates the average runtime per generated sample.}
      \label{tab:ablation_inference_steps}
    \end{table}
}

The results demonstrate that VoiceCloak exhibits remarkable stability, maintaining high defense effectiveness across a wide range of attacker inference steps. While minor fluctuations exist, key metrics like DSR and ASV remain consistently strong, indicating that the core defense is not compromised by the length of the synthesis process.
This stability stems directly from our design choice to concentrate adversarial optimization on the initial denoising timesteps ($T_{adv}=6$) that early steps are critical for reconstructing the fundamental low-frequency and coarse structural components of the audio signal. By introducing potent disruptions at this foundational stage, the errors are irrevocably embedded into the generation process.
Consequently, any subsequent refinement steps operate on an already corrupted foundation. This is evidenced by the trend in the DTW score, which increases with $T$. This suggests that as the model performs more refinement steps, the initial error is not corrected but rather cumulatively amplified.

\subsubsection{Ablation on PGD Iterations}
We conducted an ablation study to analyze the impact of the number of PGD optimization iterations on both defense effectiveness and computational efficiency. Table \ref{tab:ablation_attack_iters} presents the results as we vary the number of iterations from 5 to 100.
{
    \begin{table}[!ht]
      \centering
        \setlength{\heavyrulewidth}{1.17pt} 
        \resizebox{0.98\linewidth}{!}{
          \begin{tabular}{c|c|ccc} 
            \toprule
            \textbf{PGD Iters} & \textbf{Avg Time/sample(s)} & \textbf{ASV}($\downarrow$) & \textbf{NISQA}($\downarrow$) & \textbf{DSR}($\uparrow$) \\
            \hline
            
            5   & 15.02  & 0.178 & 2.63 & 0.564 \\
            10  & 29.81  & 0.142 & 2.46 & 0.636 \\
            25  & 74.94  & 0.110 & 2.32 & 0.734 \\
            50  & 148.66 & 0.128 & 2.33 & 0.716 \\
            100 & 295.80 & 0.100 & 2.26 & 0.720 \\
            \bottomrule
          \end{tabular}
      }
      
      \caption{Ablation study on the number of adversarial attack iterations. Avg Time is measured in seconds per sample.}
      \label{tab:ablation_attack_iters}
    \end{table}
}

The results reveal a clear relationship between the number of iterations and the performance. As expected, the average time cost increases linearly with the number of PGD iterations. In terms of defense effectiveness, we observe that performance, particularly the Defense Success Rate (DSR), generally improves as the number of iterations increases from 5 to 25.
However, \textbf{a notable point of diminishing returns} is observed beyond this. The DSR peaks at 73.4\% with 25 iterations and then slightly plateaus or fluctuates around a similar level for 50 and 100 iterations (71.6\% and 72.0\% respectively). This indicates that while the time cost continues to grow linearly, the defense performance \textbf{reaches a bottleneck after approximately 25-50 iterations}.
This analysis highlights the flexibility of VoiceCloak and informs the optimal choice of parameters based on different application needs:
\begin{itemize}
    \item \textbf{For security-critical scenarios} where maximizing defense is paramount, setting the iteration count to 50 provides a near-optimal level of protection, as further increases yield minimal performance gains at a significant additional time cost.
    \item \textbf{For efficiency-critical scenarios} where faster processing is required, setting the iteration count to 25 offers an excellent trade-off, providing strong defense effectiveness at roughly 1min.
\end{itemize}
\section{Additional Related Work}
\subsection{Adversarial Examples for Classifiers}
The paradigm of adversarial examples was first established in the domain of image classification \cite{advsamples_1,advsamples_2,advsamples_3}. The core objective is to generate an input, $x'$, that is perceptually similar to a benign input $x$ but causes a trained classifier, $f(\cdot)$, to produce an incorrect prediction. Formally, an adversarial example $x'$ is defined by two primary properties: 1. \textbf{Imperceptibility}: The perturbation between $x$ and $x'$ is minimal, such that $\mathcal{D}(x,x')\leq \epsilon$ for a given distance metric $\mathcal{D}$ and a small buget $\epsilon$. Consistent with prior work, the $l_{\infty}$-norm is a prevalent metric for this constraint. 2. \textbf{Effectiveness}: The model's prediction for the adversarial example is incorrect, i.e., $f(x') \neq y$, where $y$ is the ground-truth label for $x$.

This is typically framed as a constrained optimization problem, where the goal is to find a perturbation $\delta$ that maximizes a loss function $\mathcal{L}$ encouraging misclassification, subject to the imperceptibility constraint: $$\delta_{adv}= \arg \max_{||\delta||_p \leq \epsilon} \mathcal{L}(f(x+\delta),y).$$
This problem is commonly solved using iterative methods, with Projected Gradient Descent (PGD) \cite{PGD} being a standard and robust algorithm. PGD iteratively updates the adversarial example by taking a small step in the direction of the loss gradient and then projecting the result back onto the $\epsilon$-ball to satisfy the constraint: 
\begin{equation}
    x_{r+1} = \Pi_{\mathbf{x},\epsilon}(\mathbf{x}_r + \alpha \cdot \operatorname{sign}(\nabla_{\mathbf{x}}\mathcal{L}((f(\mathbf{x}_r), y))),
\end{equation}
where $r$ denotes the iteration, $\alpha$ is the projection operator ensuring that the updated example $\mathbf{x}_{r+1}$ remains within the $\epsilon$-ball around the original input $\mathbf{x}$.

\subsection{Adversarial Attack for Deepfake}
The principles of adversarial examples have been extended from classifiers to generative models, such as those used for Deepfake synthesis. In this domain, the attack objective shifts from causing misclassification to degrading the quality of the generated output or disrupting the synthesis process entirely.
One primary strategy involves attacking ancillary components within the generation pipeline, such as facial landmark detectors or feature extractors, which are often essential pre-processing steps for deepfake models. By corrupting the inputs to these upstream modules, the final output of the core generative model is consequently compromised. This approach is analogous to early proactive audio defenses that targeted speaker encoders \cite{attack-vc}.
Another more direct strategy targets the core generative model itself. While initial efforts focused on exploiting architectural properties of GANs to degrade image quality, these methods are often ineffective against the distinct mechanisms of Diffusion Models (DMs). For DMs, an effective adversarial example is one that the model considers "out-of-distribution", thereby hindering its superior reconstruction capabilities.
Our work builds upon this more direct approach. However, it is specifically tailored to the unique vulnerabilities of DMs (e.g., attention, score function, U-Net features), addressing the limitations of defenses designed for prior generative paradigms.

\putbib[appendix]  
\end{bibunit}
\end{document}